

\input harvmac

\overfullrule=0pt


\def\C{{\scriptscriptstyle C}}

\def\J{{\scriptscriptstyle J}}

\def\L{{\scriptscriptstyle L}}

\def\Q{{\scriptscriptstyle Q}}

\def\S{{\scriptscriptstyle S}}

\def\W{{\scriptscriptstyle W}}

\def\Z{{\scriptscriptstyle Z}}


\def\CA{{\cal A}}

\def\CM{{\cal M}}

\def\CO{{\cal O}}


\def\d{\delta}
\def\e{\epsilon}

\def\s{\sigma}
\def\t{\tau}
\def\th{\theta}
\def\u{\mu}
\def\v{\nu}


\def\aEM{\alpha_{\scriptscriptstyle EM}}
\def\aS{\alpha_s}

\def\bbar{{\overline b}}
\def\bbbaroctet{b\bbar[^3S_1^{(8)}]}
\def\bbbarsinglet{b\bbar[^3S_1^{(1)}]}

\def\Br{{\rm Br}}
\def\cbar{{\overline c}}
\def\ccbaroctet{c\cbar[^3S_1^{(8)}]}
\def\ccbarsinglet{c\cbar[^3S_1^{(1)}]}
\def\ccdot{\hbox{\kern-.1em$\cdot$\kern-.1em}}
\def\chiQJ{\chi_{\Q\J}}

\def\decaysto{\raise1.1ex\hbox{$\lfloor$\kern-0.25em\lower1.125ex
  \hbox{$\longrightarrow\>\,$}}}	
\def\dilog{{\rm Li_2}}

\def\GeV{\>\, \rm GeV}

\def\gtap{\raise.3ex\hbox{$>$\kern-.75em\lower1ex\hbox{$\sim$}}}
\def\Jpsi{J/\psi}
\def\long{{\rm long}}
\def\LQCD{\Lambda_{\scriptscriptstyle QCD}}
\def\ltap{\raise.3ex\hbox{$<$\kern-.75em\lower1ex\hbox{$\sim$}}}
\def\Mb{M_b}
\def\Mc{M_c}

\def\MQ{M_\Q}

\def\MZ{M_\Z}
\def\Nc{N_c}

\def\octetelement{\CM_8(\psi_Q)}
\def\pbar{{\overline{p}}}

\def\pperp{p_\perp}
\def\prompt{{\rm prompt}}
\def\psiQ{\psi_\Q}

\def\pT{p_\perp}
\def\qbar{{\overline q}}
\def\Qbar{{\overline Q}}

\def\QQbaroctet{Q\Qbar[^3S_1^{(8)}]}

\def\QQbarsinglet{Q\Qbar[^3S_1^{(1)}]}

\def\short{{\rm short}}
\def\singletelement{\CM_1(\psi_Q)}

\def\thW{{\th_\W}}
\def\total{{\rm total}}



\newdimen\pmboffset
\pmboffset 0.022em
\def\oldpmb#1{\setbox0=\hbox{#1}%
 \copy0\kern-\wd0 \kern\pmboffset\raise
 1.732\pmboffset\copy0\kern-\wd0 \kern\pmboffset\box0}


%
%
\def\appendix#1#2{\global\meqno=1\global\subsecno=0\xdef\secsym{\hbox{#1.}}
\bigbreak\bigskip\noindent{\bf Appendix. #2}\message{(#1. #2)}
\writetoca{Appendix {#1.} {#2}}\par\nobreak\medskip\nobreak}


\nref\Opalups{The OPAL collaboration, ``Observation of $\Upsilon$
  production in hadronic $Z^0$ decays, OPAL Physics note PN192, (1995),
  unpublished.}
\nref\Opalpsi{The OPAL collaboration, ``$\Jpsi$ and $\psi'$ production in
  hadronic $Z^0$ decays'', OPAL Physics Note PN178 (1995), unpublished.}
\nref\Delphisearch{The DELPHI collaboration, ``Search for promptly produced
  heavy quarkonium states in hadronic $Z$ decays'', CERN-PPE/Paper 0116
 (1995), unpublished.}
\nref\Delphipsi{The DELPHI collaboration, Phys. Lett. {\bf B341} (1994) 109.}
\nref\BCY{E. Braaten, K. Cheung and T.C. Yuan, Phys. Rev. {\bf D48} (1993)
 4230.}
\nref\Hagiwara{K. Hagiwara, A.D. Martin and W.J. Stirling, Phys. Lett. {\bf
B267} (1991) 527\semi (E) Phys. Lett. {\bf B316} (1993) 631.}
\nref\BraatenYuan{E. Braaten and T.C. Yuan, Phys. Rev. {\bf D50} (1994)
 3176.}
\nref\BDFM{E. Braaten, M.A. Doncheski, S. Fleming and M.L. Mangano,
 Phys. Lett. {\bf B333} (1994) 548.}
\nref\Roy{D.P. Roy and K. Sridhar, Phys. Lett. {\bf B339} (1994) 141.}
\nref\Cacciari{M. Cacciari and M. Greco, Phys. Rev. Lett. {\bf 73} (1994)
 1586.}
\nref\ChoWise{P.Cho and M. Wise, Phys. Lett. {\bf B346} (1995) 129.}
\nref\BraatenFleming{E. Braaten and S. Fleming, Phys. Rev. Lett. {\bf 74}
 (1995) 3327.}
\nref\ChoLeibov{P. Cho and A.K. Leibovich, CALT-68-1988 (1995),
  unpublished.}
\nref\CGMP{M. Cacciari, M. Greco, M.L. Mangano and A. Petrelli,
 CERN-TH/95-129 (1995), unpublished.}
\nref\Bodwin{G.T. Bodwin, E. Braaten and G.P. Lepage, Phys. Rev. {\bf D51}
(1995) 1125.}
\nref\Kuhn{J. H. K\"uhn, J. Kaplan and E. G. O. Safiani, Nucl. Phys. {\bf
 B157} (1979) 125.}
\nref\Guberina{B. Guberina, J.H. K\"uhn, R.D. Peccei and R. R\"uckl, Nucl.
 Phys. {\bf B174} (1980) 317.}
\nref\BraatenYuanI{E. Braaten and T.C. Yuan, Phys. Rev. Lett. {\bf 71} (1993)
 1673.}
\nref\Quigg{E.J. Eichten and C. Quigg, Phys. Rev. {\bf D47} (1995) 1726.}
\nref\CDFPsidata{The CDF collaboration, Fermilab-Conf-94/136-E (1994),
 unpublished.}
\nref\Papadimitriou{The CDF collaboration, Fermilab-Conf-95/128-E (1995),
 unpublished.}
\nref\CDFUpsilondata{The CDF collaboration, Fermilab-Conf-94/221-E (1994),
 unpublished.}
\nref\Lthreepsi{The L3 collaboration, Phys. Lett. {\bf B288} (1992) 412.}
\nref\CKY{K. Cheung, W.-Y. Keung and T.C. Yuan, Fermilab-Pub-95/300-T (1995),
 unpublished.}


\nfig\nonfragfig{Feynman diagrams which mediate the lowest order
color-octet process \break $Z \to g + Q \Qbar[{}^{2\S+1} L_\J^{(8)}]
\to \psiQ + X$.}
\nfig\Qfragfig{Feynman diagrams which mediate the heavy quark fragmentation
process \break
$Z \to Q + \Qbar + Q \Qbar[ ^{2\S+1} L_\J^{(8)}] \to \psiQ + X$.  Two other
non-fragmentation graphs whose $O(\aS^2)$ contributions are subleading in
axial gauge are not pictured.}
\nfig\gluefragfig{Feynman diagrams which mediate the gluon fragmentation
process $Z \to q + \qbar + g^* \break \to  q + \qbar + \QQbaroctet \to
\psiQ + X$.}
%


\def\CITTitle#1#2#3{\nopagenumbers\abstractfont
\hsize=\hstitle\rightline{#1}
\vskip 0.4in\centerline{\titlefont #2} \centerline{\titlefont #3}
\abstractfont\vskip .4in\pageno=0}

\CITTitle{{\baselineskip=12pt plus 1pt minus 1pt
  \vbox{\hbox{CALT-68-2020}\hbox{DOE RESEARCH AND}\hbox{DEVELOPMENT
  REPORT}}}}
{Prompt Upsilon and Psi Production at LEP}{}
\centerline{
  Peter Cho\footnote{$^1$}{Work supported in part by a DuBridge Fellowship and
  by the U.S. Dept. of Energy under DOE Grant no. DE-FG03-92-ER40701.}}
\centerline{Lauritsen Laboratory}
\centerline{California Institute of Technology}
\centerline{Pasadena, CA  91125}

\vskip .2in
\centerline{\bf Abstract}
\bigskip

	Color-octet contributions to quarkonia production at LEP are studied
herein.  The short distance formation of heavy quark-antiquark pairs in
color-octet configurations via gluon fragmentation processes is significantly
enhanced relative to the creation of color-singlet pairs via heavy quark
fragmentation.  But the subsequent long distance hadronization of these
colored pairs into physical quarkonium bound states is suppressed compared to
the nonperturbative evolution of their colorless counterparts.  We find that
the overall LEP rates for gluon fragmentation into prompt Upsilon and Psi
vector bosons exceed those from heavy quark fragmentation.  Inclusion of the
dominant color-octet quarkonium production channel eliminates sizable
discrepancies between previous predictions and recent measurements of prompt
$Z \to \Jpsi+X$, $Z \to \psi'+X$ and $Z \to \Upsilon+X$ branching fractions.

\Date{9/95}


\newsec{Introduction}

	Rare decays of $Z$ bosons to heavy quarkonia final states are
currently under study at LEP \refs{\Opalups{--}\Delphipsi}.  The
predominant majority of charmonia observed at the electron-positron facility
originate from $B$ meson decays.  A smaller fraction come from prompt
processes which create charm-anticharm pairs at the primary interaction
point.  These two qualitatively different sources of $c\cbar$ mesons can
be distinguished via vertex detection measurements.  In the $b\bbar$ sector,
no delayed production mechanism exists, for bottomonia are among the heaviest
hadrons created in $Z$ decays.  Bottomonia consequently result only from
prompt reactions.

	During the past few years, a number of theoretical predictions for
quarkonia production rates at LEP have appeared in the literature.  The heavy
quark fragmentation processes
\eqn\Qfrag{\eqalign{Z \to \> & Q + \Qbar\cr & \, \decaysto \psiQ +Q \cr}
\quad\qquad {\rm and} \quad\qquad
\eqalign{Z \to Q + &\Qbar\cr & \, \decaysto \psiQ + \Qbar \cr}}
have been claimed to represent the main source of prompt Upsilon and Psi
vector bosons \BCY.
The rates for these modes are orders of magnitude larger than those for
non-fragmentation channels such as the gluon radiation reaction
$Z \to \psiQ g g$.  But the recent first detection of Upsilons by the
OPAL collaboration casts doubt upon the assertion that the decay processes
in \Qfrag\ account for most bottomonia produced at LEP \Opalups.  In
particular, the measured branching ratio
\eqn\UpsBr{\sum_{n=1}^3 \Br\bigl(Z \to \Upsilon(nS)+ X\bigr) =
\bigl( 1.2^{\displaystyle{+0.9}}_{\displaystyle{-0.6}} \pm 0.2 \bigr)
\times 10^{-4}}
is ten times larger than the prediction based upon bottom quark fragmentation.
Similarly, the numbers of prompt $\Jpsi$ and $\psi'$ mesons which OPAL has
observed \Opalpsi
\eqn\PsiBr{\eqalign{\Br(Z \to \Jpsi + X)_\prompt &=
{N(\Jpsi)_\prompt \over N(\Jpsi)_\total} \times \Br(Z \to \Jpsi+X)_\total \cr
&= (0.077 \pm 0.058) \times (3.9 \pm 0.2 \pm 0.3) \times 10^{-3} =
(3.0 \pm 2.3) \times 10^{-4} \cr
& \cr
\Br(Z \to \psi'+X)_\prompt &=
{N(\psi')_\prompt \over N(\psi')_\total} \times \Br(Z \to \psi'+X)_\total \cr
&= (0.14 \pm 0.09) \times (1.6 \pm 0.3 \pm 0.2) \times 10^{-3} =
(2.2 \pm 1.5) \times 10^{-4} \cr}}
exceed the abundance expected from the modes in \Qfrag.  Although these OPAL
findings are preliminary and have yet to be confirmed by other LEP groups,
the large discrepancies between the data and existing theoretical estimates
are interesting.

	In this article, we investigate a potential resolution to this
puzzle.  We suggest that the largest contribution to prompt quarkonia
production stems from gluon rather than heavy quark fragmentation.
This possibility was first considered in 1991 by Hagiwara, Martin and Stirling
\Hagiwara.  The branching ratio calculated by these authors for the
gluonic process
\eqn\gluefrag{\eqalign{Z \to q + \qbar + & g^* \cr
                       & \decaysto \psiQ + X \cr}}
turned out to be smaller than the branching fraction found later for
the quark mode in \Qfrag.  But their result was based upon the
assumption that the entire reaction in \gluefrag\ takes place at an energy
scale greater than or equal to the $\psiQ$ mass.  Since
$Z \to q\qbar g^* \to q\qbar\psiQ g g$ decay starts at $O(\aS^4)$, it
is not surprising that the rate for the purely short distance part of this
transition is small.

	Theoretical understanding of gluon fragmentation into heavy
quarkonia has significantly matured since the time Hagiwara
{\it et al.} carried out their analysis.  High energy formation of
heavy quark-antiquark pairs in color-octet configurations followed by low
energy evolution into color-singlet hadrons is now known to represent the
dominant source of large $\pT$ quarkonia at hadron colliders
\refs{\BraatenYuan{--}\BraatenFleming}.  Inclusion of gluon
fragmentation mechanisms into charmonia and bottomonia cross section
calculations appears to eliminate orders of magnitude discrepancies between
theoretical predictions and recent Tevatron measurements
\refs{\ChoLeibov,\CGMP}.  However, it is important to test this new
understanding of quarkonia production in other settings.  We are thus
motivated to study the implications of color-octet mechanism ideas
for $Z$ decays to Upsilon and Psi final states.

	Our paper is organized as follows.  We first calculate the LEP
rates for charmonia and bottomonia production arising from lowest order
color-octet processes in section 2.  We then review the heavy quark
fragmentation prediction for both color-singlet and color-octet quarkonia
production in section 3. We examine next in section 4 the gluon fragmentation
contribution to $Z \to \Upsilon+X$ and $Z \to \psi+X$.  Finally, we compare
all these different production modes in section 5 and close with a summary
of our findings.

\newsec{Lowest order color-octet quarkonia production}

	Quarkonia production at LEP involves several different length
scales.  The first is set by the $Z$ boson mass $\MZ$ which represents the
$e^+ e^-$ collider's current operating energy.  The second is fixed by
the mass $\MQ$ of the quark or antiquark inside a quarkonium state.  The
heavy constituents are bound together over a distance scale determined by
their momentum $\MQ v$, and they interact with each other over time periods
set by their kinetic energy $\MQ v^2$.  Since the velocity $v \sim 1/\log \MQ$
of the $Q$ and $\Qbar$ inside the bound state is small
compared to the speed of light, these scales are all widely separated from
one another:
\eqn\scales{ (\MQ v^2)^2 \ll (\MQ v)^2 \ll \MQ^2 \ll \MZ^2.}

	The presence of so many different energy scales complicates the
analysis of quarkonium production.  But the most important characteristics
may be simply distilled within an effective field theory framework called
Nonrelativistic Quantum Chromodynamics (NRQCD) which systematically keeps
track of the hierarchy in \scales\ \Bodwin.  This effective theory is
based upon a double power series expansion in the short distance strong
interaction fine structure constant $\aS$ and the small velocity parameter
$v$.  We will work throughout this paper within the NRQCD framework.

	To begin, we consider the decay mode
\eqn\ZgQQbardecay{Z \to g + Q \Qbar[ ^{2\S+1} L_\J^{(8)}] }
which creates a colored heavy quark-antiquark pair recoiling against a hard
gluon.  The lowest order diagrams that mediate this transition are
illustrated in \nonfragfig.  The spin, orbital and total angular momentum
quantum numbers of the $Q\Qbar$ pair are indicated in spectroscopic
notation inside the square brackets in \ZgQQbardecay, and its color is
labeled by the octet superscript.  Using the projection techniques described
in refs.~\refs{\ChoLeibov,\Kuhn,\Guberina}, we can start with the general
$Z \to g + Q + \Qbar$ amplitude and pick out terms which correspond to the
formation of a $Q\Qbar$ pair in a specified partial wave, spin and color
state.  For example, the short distance production amplitude for a colored
$L=0$, $S=1$ pair is given by
\eqn\ZgQQbaramp{\eqalign{
i \CA\bigr( Z(k+P) \to g_a(k) + & \QQbaroctet_b (P) \bigl)_\short = \cr
& {2 g_2 g_3 \over \cos\thW} {M \over (\MZ^2 - M^2)} T_3 \d_{ab}
\e^{\u\v\s\t} \varepsilon_\u(k)^* \varepsilon_\v(P)^* \varepsilon_\s(k+P)
k_\t. \cr}}
Here $M=2\MQ$ represents the pair's mass, $g_2$ and $g_3$ denote the
$SU(2)_\L$ and $SU(3)_\C$ gauge couplings, $T_3$ stands for the third weak
isospin eigenvalue and $\thW$ equals the weak mixing angle.  The outgoing
$Q$ and $\Qbar$ in \nonfragfig\ propagate nearly on-shell with a combined
momentum $P$.  But the intermediate heavy quark or antiquark line
which attaches to the decaying $Z$ is off-shell by an amount of $O(\MZ^2)$.
As we shall see, the intermediate particle's extreme virtuality strongly
suppresses the lowest order color-octet process in \ZgQQbardecay\ relative
to quark and gluon fragmentation reactions that take place at higher order
in the $\aS$ expansion.

	After the colored $Q\Qbar[ ^{2\S+1} L_\J^{(8)}]$ pair is created
within a spacetime interval set by the $Z$ boson mass, it evolves over a
much longer time and distance scale into a physical quarkonium bound state.
The pair transforms into a colorless hadron via the emission or absorption
of one or more soft gluons.  These long wavelength gluons bleed off the
color of the $Q\Qbar[ ^{2\S+1} L_\J^{(8)}]$ object, but they carry
away only $O(\MQ v^2)$ worth of energy and momentum \Bodwin.  In the
$\MQ \to \infty$ limit, this hadronization process should be
perturbatively computable.  However in the real world where $\MQ
v^2 \simeq \LQCD$, the amplitude for a heavy color-octet pair to evolve
into a physical quarkonium state is nonperturbative.  The long distance
matrix element therefore cannot be computed from first principles and must
instead be determined from experiment.

	Combining together the square of the short distance amplitude
in \ZgQQbaramp\ with the long distance factor $\CM_8(\psiQ) \equiv
|\CA\bigl(\QQbaroctet \to \psiQ(n)\bigr)_\long |^2$, we obtain the partial
width
\eqn\nonfragrate{\Gamma(Z \to g + \QQbaroctet \to \psiQ + X) = {\pi\over 3}
{\aEM \aS \over \sin^2\thW \cos^2\thW } \Bigl({\Nc^2-1 \over 2} \Bigr)
\Bigl(1- {M^4 \over \MZ^4} \Bigr) {\CM_8(\psiQ) \over \MZ}}
which is averaged (summed) over initial (final) polarizations and colors.
The color-octet squared amplitude decomposes into an infinite sum
over NRQCD matrix elements
\eqn\octetdecomp{\eqalign{\CM_8 \bigl(\psi_Q(n)\bigr) = {1 \over 24 \MQ}
\sum_{m \ge n} \Bigl\{ & \langle 0 | \CO_8^{\psiQ(m)}(^3S_1) | 0 \rangle \;
\Br \bigl(\psiQ(m) \to \psiQ(n) \bigr)\cr
& + \sum_{J=0,1,2} \langle 0 | \CO_8^{\chiQJ(m)}(^3S_1) | 0 \rangle \;
 \Br\bigl(\chiQJ(m) \to \psiQ(n) \gamma \bigr) + \cdots \Bigr\} \cr}}
which determine the probabilities for a $\QQbaroctet$ pair to turn
into a $\psiQ$ through all possible intermediate channels.  Numerical values
for these NRQCD matrix elements have recently been extracted from CDF
Upsilon and Psi cross section measurements in refs.~\refs{\ChoLeibov,\CGMP}.
Once the universal matrix elements are known from one experiment, they may be
applied to the study of color-octet quarkonia production at any other.

	Inserting the long distance squared amplitude values
\eqn\Moctetvalues{\eqalign{\CM_8(\Jpsi) &= 6.8 \times 10^{-4} \GeV^2 \cr
\CM_8(\psi') & = 2.0 \times 10^{-4} \GeV^2 \cr
\sum_{n=1}^3 \CM_8(\Upsilon(nS)) &= 6.4 \times 10^{-3} \GeV^2 \cr}}
into \nonfragrate\ along with the parameters $\Mc = 1.48 \GeV$,
$\Mb = 4.88 \GeV$, $\aS(2 \Mc) = 0.28$ and $\aS(2 \Mb) = 0.19$, we find the
$Z \to \Jpsi + X$, $Z \to \psi' + X$, and
$Z \to \sum_n \Upsilon(nS) + X$ branching fractions listed in the first row
of Table~I.  Comparing these predictions with the OPAL numbers in
eqns.~\UpsBr\ and \PsiBr, we see that the contribution to
Upsilon and Psi production at LEP from the color-octet process in
\ZgQQbardecay\ with an $L=0$, $S=1$ intermediate state is negligible.
Yields from similar reactions involving colored $Q\Qbar$ pairs with different
orbital and spin angular momenta should be even smaller since their
hadronization into physical bound states takes place at higher orders in the
velocity expansion.  We therefore conclude that $O(\aS)$ color-octet
quarkonia production channels can be ignored without loss.

\newsec{Heavy quark fragmentation}

	Parton fragmentation represents the dominant source of heavy
quarkonia in high energy collisions \BraatenYuanI.  As we have seen,
the propagators of short distance intermediate partons
are usually off-shell by $O(1/E^2)$ where $E$ denotes some characteristic
energy of the hard reaction.  But in fragmentation
processes, the quark or gluon which splits into a quarkonium state plus
other partons is typically off-shell by only $O(1/\MQ^2)$.  At sufficiently
high energies, fragmentation contributions to quarkonia cross sections
dominate over all non-fragmentation competitors since the former
are enhanced relative to the latter by powers of $E^2/\MQ^2$.  This outcome
is guaranteed to occur even if the short distance part of the fragmentation
reaction takes place at higher order in the $\aS$ expansion, for additional
factors of the QCD coupling only logarithmically suppress reaction rates with
increasing $E$.

	One of the first applications of these fragmentation ideas was to
charmonia and bottomonia production at LEP.  In ref.~\BCY, Braaten, Cheung
and Yuan calculated the probabilities for charm and bottom quarks to
respectively split into Psi and Upsilon bound states.  These authors then
folded together the fragmentation probabilities with the $Z \to c\cbar$ and
$Z \to b\bbar$ partial widths and determined the prompt quarkonia branching
fraction $\Br(Z \to \psiQ + X)_{Q \> {\rm frag}}$.  We will review the
findings of Braaten {\it et al.} below and then generalize their color-singlet
results to include color-octet contributions as well.

	Heavy quark fragmentation production of Upsilon and Psi bosons at LEP
proceeds at $O(\aS^2)$ through the Feynman graphs illustrated in \Qfragfig.
\foot{The two diagrams displayed in \Qfragfig\ dominate in axial gauge over
two other $Z \to \psiQ Q \Qbar$ graphs which do not possess an obvious
fragmentation interpretation.  Analysis of $\psiQ$ production via heavy quark
fragmentation is most transparent in this particular gauge.}
The mass $s$ of the virtual heavy quark or antiquark which
splits into the outgoing $Q\Qbar$ pair ranges over the interval $3 \MQ
\le s \le \MZ-\MQ$.  $\psiQ$ production is maximized when the pair is created
near the $\MQ$ scale in a color-singlet $L=0$, $S=1$ configuration.  In the
fragmentation approximation, the contribution of this channel to the
$Z \to \psiQ$ partial width simply equals the product of the $Z \to Q \Qbar$
decay rate and the hadronization probabilities $D_{Q \to \psiQ}$ and
$D_{\Qbar \to \psiQ}$ \BCY:
\eqn\Qfragrate{\Gamma(Z \to Q + \Qbar + \QQbarsinglet \to \psiQ + X) =
\Gamma(Z \to Q \Qbar) \times \bigl( D_{Q \to \psiQ} + D_{\Qbar \to \psiQ}
\bigr)}
where
\eqn\Qfragprobs{D_{Q \to \psiQ} = D_{\Qbar \to \psiQ} =
{8 \aS(M)^2 \over \Nc} \Bigl( {\Nc^2-1 \over 2 \Nc} \Bigr)^2
\Bigl[ {1189 \over 30} - 57 \ln 2 \Bigr] {\CM_1(\psiQ) \over M^2}.}
All nonperturbative information associated with $\psiQ$ bound state formation
resides within the squared amplitude
\eqn\singletdecomp{\eqalign{\CM_1 \bigl(\psi_Q(n)\bigr) &=
|\CA\bigl(\QQbarsinglet \to \psiQ(n)\bigr)_\long |^2 \cr
&= {1 \over 6 \Nc \MQ} \sum_{m \ge n}
\Bigl\{ \langle 0 | \CO_1^{\psiQ(m)}(^3S_1) | 0 \rangle \;
\Br \bigl(\psiQ(m) \to \psiQ(n) \bigr) + \cdots \Bigr\}. \cr}}
Numerical values for the color-singlet NRQCD matrix elements which enter at
lowest order in the velocity expansion into this long distance factor can be
extracted from the charmonia and bottomonia wavefunctions at the origin
tabulated in ref.~\Quigg.  Adopting the values
\eqn\Msingletvalues{\eqalign{
\CM_1(\Jpsi) &= 6.1 \times 10^{-2} \GeV^2 \cr
\CM_1(\psi') &= 2.9 \times 10^{-2} \GeV^2 \cr
\sum_{n=1}^3 \CM_1(\Upsilon(nS)) &= 2.3 \times 10^{-1} \GeV^2, \cr}}
we derive the heavy quark fragmentation branching ratios listed in the
second row of Table~I.

	It is instructive to compare these color-singlet results with their
color-octet analogues.  The partial width for $Z \to Q+\Qbar+\QQbaroctet
\to \psiQ + X$ is readily obtained from $\Gamma(Z \to Q+\Qbar+\QQbarsinglet
\to \psiQ+X)$ by performing the following color-factor and squared amplitude
substitutions in eqn.~\Qfragprobs:
\eqn\substitutions{ \Bigl({\Nc^2-1 \over 2 \Nc}\Bigr)^2 \to {\Nc^2-1 \over
4 \Nc^2} \qquad {\rm and} \qquad \CM_1(\psiQ) \to \CM_8(\psiQ).}
These alterations diminish both the short and long distance contributions
to the color-octet rate relative to its color-singlet counterpart.  The
color-octet heavy quark fragmentation branching ratios appearing in the third
row of Table~I are consequently much smaller than their color-singlet
counterparts.

	Comparing the predicted rates for the processes in
eqn.~\Qfrag\ with the Upsilon and Psi measurements in eqns.~\UpsBr\ and
\PsiBr, we see that heavy quark fragmentation does not appear to account for
most prompt quarkonia observed at LEP.  Although the data are preliminary and
their error bars are large, the sizable mismatch between theory and
experiment suggests that other rapid production mechanisms are at work.
We therefore turn to consider gluon fragmentation contributions to Upsilon
and Psi production in the following section.

\newsec{Gluon fragmentation}

	A number of striking disagreements between theoretical predictions
and experimental measurements of quarkonia cross sections at hadron colliders
have arisen during the past few years.  The CDF collaboration has discovered
that prompt $\Jpsi$ and $\psi'$ production at the Tevatron exceed
color-singlet differential rate predictions by orders of magnitude
\refs{\CDFPsidata,\Papadimitriou}.  CDF has also observed greater numbers of
$\Upsilon(1S)$, $\Upsilon(2S)$ and $\Upsilon(3S)$ vector mesons than
were originally anticipated \CDFUpsilondata.  These experimental surprises
from Fermilab have stimulated several recent theoretical investigations of
quarkonia creation in general and gluon fragmentation processes in particular
\refs{\BraatenYuan{--}\CGMP}.  The incorporation of color-octet production
mechanisms into cross section calculations has been demonstrated to yield
significantly better descriptions of charmonia and bottomonia measurements
than those based upon color-singlet mechanisms alone.  Since numerical values
for long distance color-octet matrix elements are {\it a priori} unknown, the
absolute magnitude for $d\sigma(p\pbar \to \psiQ+X)/d\pperp$ must simply be
fitted to data.  But the shapes of the predicted and measured differential
distributions at high $\pperp$ agree quite well once gluon fragmentation
reactions are taken into account.  Moreover, the fitted values for the
color-octet matrix elements are consistent with NRQCD velocity counting rules.
Color-octet contributions thus appear to eliminate large discrepancies
between theory and experiment.

	Given that gluon fragmentation plays a central role in
quarkonia production at the Tevatron, it is interesting to study
its impact at LEP as well.  Gluon fragmentation mediated decays of $Z$ bosons
into Upsilon and Psi final states start at $O(\aS^2)$ through the reaction
displayed in eqn.~\gluefrag\ and illustrated in \gluefragfig.  A
straightforward computation yields the differential rate for this process:
\eqn\diffrate{\eqalign{
{d^2 \Gamma \over d E_1 d E_2}(Z & \to q + \qbar + g^* \to \psiQ+X) =
{16 K \over \d^2} \Bigl\{ 2(E_1^2+E_2^2) (1-2 E_1)(1-2 E_2) \cr
& \qquad + \d^2 \bigl[ 8 E_1 E_2(E_1+E_2) - 2(3 E_1^2 + 4 E_1 E_2 + 3 E_2^2)
  + 4 (E_1 + E_2) - 1 \bigr] \cr
& \qquad + \d^4 (1-2 E_1)(1-2 E_2) \Bigr\} / \bigl[ (1-2 E_1)^2 (1-2 E_2)^2
\bigr]. \cr}}
Here $E_1$ and $E_2$ represent the dimensionless lab frame energies of the
quark and antiquark scaled relative to the $Z$ mass, $\d = M/\MZ$ equals
the rescaled quarkonium mass and $K$ denotes a constant prefactor
\eqn\prefactor{K = \aS^2 \Bigl( {\Nc^2-1 \over 2\Nc} \Bigr) \Gamma(Z \to
q\qbar) {\octetelement \over \MZ^2}}
which contains the rescaled long distance matrix element
$\CM_8(\psiQ)/\MZ^2$.  The $O(\d^2)$ and $O(\d^4)$ terms inside the curly
brackets in \diffrate\ are negligibly small compared to the leading $O(\d^0)$
term since $\d^2 \ll 1$.  Similarly, corrections to the $Z \to q\qbar g^*$
partial width from the masses of the $q$ and $\qbar$ are unimportant and have
been ignored.

	Integrating the differential expression in \diffrate\ over the quark
and antiquark energy limits $0 \le E_1 \le (1-\d^2)/2$ and
$(1-\d^2)/2 - E_1 \le E_2 \le [1-\d^2/(1-2 E_1)]/2$, we obtain the partial
decay rate \Hagiwara
\eqn\gfragrate{\eqalign{\Gamma(Z \to q + \qbar + g^* \to \psiQ+X) &=
{2 K \over \d^2} \Bigl\{ (1+\d^2)^2 \log \d^2 \log{ \d^2 \over (1+\d^2)^2}
+ (3+4\d^2+3 \d^4) \log \d^2 \cr
& + 5(1-\d^4) + 2 (1+\d^2)^2 \Bigl[ \dilog \Bigl( {\d^2 \over 1+\d^2}
\Bigr) - \dilog \Bigl( {1 \over 1+\d^2} \Bigr) \Bigr] \Bigr\}. \cr}}
The $1/\d^2$ term appearing alongside the constant prefactor $K$ originates
from the propagator
of the virtual gluon.  We again recall that the fragmenting $g^*$ is off-shell
by an amount of order $\MQ^2$ rather than $\MZ^2$.  The integrated rate
for the gluonic process is thus enhanced by the same $1/\d^2$
factor as its heavy quark fragmentation analogue.

	The gluonic channel's rate is further amplified by a large double
logarithm.  The source of the $(\log \d^2)^2$ factor appearing in the first
term of \gfragrate\ can be traced to the double pole residing in the first
term of \diffrate.  The double log in the integrated rate results from the
overlap of an infrared singularity with a collinear divergence.  As
$\log \d^2$ is numerically large in both the charmonia and bottomonia sectors,
higher order corrections to the gluon fragmentation partial width are likely
to be significant.  The leading logarithms should be resummed in order to
obtain a more reliable estimate for the rate at which $\psiQ$ quarkonia are
generated via gluon fragmentation at LEP.  Performing such a resummation is
rather subtle, and we will return to this issue in a future work.  For now,
we will simply evaluate the lowest order expression with the understanding
that subleading corrections may not be negligible.

	With the extra double log enhancement factor, the short distance
contribution to the gluon fragmentation process in \gluefrag\ dominates over
that for the heavy quark fragmentation reaction in \Qfrag.  But the long
distance color-octet matrix element $\CM_8(\psiQ)$ for the former is
numerically two orders of magnitude smaller than the corresponding
color-singlet squared amplitude $\CM_1(\psiQ)$ for the latter.  The relative
importance of these two different fragmentation mechanisms thus depends upon
the outcome of a competition between short and long distance factors.  To
determine the winner, we evaluate the partial width in \gfragrate\ and list
the resulting $Z \to \Upsilon$ and $Z \to \psi$ branching fractions
in the last row of Table~I.  Comparing these entries with the others in the
table, we see that the gluon fragmentation rate beats those for all other
prompt quarkonia production channels.

	Inclusion of the dominant color-octet contributions brings theoretical
branching fraction predictions into line with the experimental values quoted
in eqns.~\UpsBr\ and \PsiBr.  These results consequently represent a nontrivial
success of the color-octet picture.  We also note that the rate for gluon
fragmentation production of $\Jpsi$ is consistent with the L3 upper bound
$\Br(Z \to q\qbar g^* \to \Jpsi+X) < 7.0 \times 10^{-4}$ \Lthreepsi\ and
the more recent DELPHI limit $\Br(Z \to q\qbar g^* \to \Jpsi+X) < 4.1
\times 10^{-4}$ \Delphisearch.  Improved LEP measurements of this branching
ratio along with its $\psi'$ and $\Upsilon$ analogues would provide useful
cross checks on NRQCD color-octet matrix element values.  Angular distribution
and spin alignment data would further test color-octet mechanism ideas.  In
short, further observation of $Z$ decays to Upsilon and Psi final states
would enhance understanding of basic quarkonium physics.

\newsec{Conclusions}

	In this article, we have investigated color-octet contributions to
prompt quarkonia production at LEP.
\foot{Similar work has recently been reported by Cheung, Keung and Yuan in
ref.~\CKY.}
The quantitative results of our study compiled in Table~I can be qualitatively
understood by considering the order of magnitude estimates displayed in
Table~II.  In the first column, we list the perturbative QCD order for each
of the production channels that we have considered in this paper.  We next
recall in the second column the short distance factors associated with each
mode.  As we have seen, the $1/\d^2=\MZ^2/M^2$ enhancement of the fragmentation
reactions overwhelms their $O(\aS)$ suppression relative to lowest order
non-fragmentation processes.  Rates for the former consequently swamp those
for the latter.  The branching fraction for the gluonic mode in \gluefrag\ is
further amplified by a large double log.  In the third and fourth columns of
Table~II, we enumerate the relevant color and flavor factors for each of the
production channels.  It is important to note that these factors enhance the
gluon fragmentation mode by an order of magnitude relative to its heavy quark
counterpart.  Finally, we list the long distance matrix elements associated
with each process.  An interesting interplay between the various factors
appearing in the columns of this table determines the relative importance of
the production mechanisms listed in the separate rows.

	The nonperturbative suppression of color-octet processes generally
implies that they are subdominant compared to color-singlet reactions.
Intermediate $\bbbaroctet$ and $\ccbaroctet$ contributions to bottom and
charm quark fragmentation into Upsilon and Psi final states are
negligible compared to those from $\bbbarsinglet$ and $\ccbarsinglet$
pairs.  But in certain circumstances, short distance enhancement of color-octet
modes can offset their long distance suppression.  A striking example of
this phenomenon has recently been proposed as a resolution to the CDF $\psi'$
surplus problem \BraatenFleming.   Short distance enhancement of $g \to
\ccbaroctet$ could yield 50 times more high $\pperp$ $\psi'$ charmonia
than $g \to \ccbarsinglet$ even though the hadronization rate for
colored pairs from the first reaction is suppressed by $O(v^4)$ compared to
that for colorless pairs from the second.  A similar phenomenon may
play an important role at LEP.  We have found that inclusion of
the dominant gluon fragmentation channel \gluefrag\ removes sizable
disparities between previous theoretical predictions and recent experimental
measurements of $Z \to \Jpsi+X$, $Z \to \psi'+X$ and $Z \to \Upsilon+X$
branching fractions.  Color-octet mechanism ideas thus appear to solve puzzles
that have been encountered in both hadron and lepton collider settings.
Our confidence is therefore bolstered that these same ideas can be applied to
quarkonia problems in other contexts as well.

\bigskip
\centerline{{\bf Acknowledgments}}
\bigskip

	It is a pleasure to thank Stan Brodsky, Howard Georgi, Ian Hinchliffe
and Mark Wise for helpful discussions.

\vfill\eject

$$ \vbox{\offinterlineskip
\def\tablerule{\noalign{\hrule}}
\hrule
\halign {\vrule#& \strut#&
\ \hfil#\hfil & \vrule#&
\ \hfil#\hfil & \vrule#&
\ \hfil#\hfil & \vrule#&
\ \hfil#\hfil & \vrule# \cr
\tablerule%
height10pt && \omit && \omit && \omit && \omit &\cr
&& Production && $\Br(Z \to \Jpsi)$ && $\Br(Z \to \psi')$ &&
${\displaystyle \sum_{n=1}^3} \Br(Z \to \Upsilon(nS))$ &\cr
&& Mechanism && && && &\cr
height10pt && \omit && \omit && \omit && \omit & \cr
\tablerule
height10pt && \omit && \omit && \omit && \omit & \cr
&& \quad $Z \to g+\QQbaroctet$ \quad &&
\quad $1.6\times10^{-7}$\quad && \quad $4.6 \times 10^{-8}$ \quad &&
\quad $1.0 \times 10^{-6}$ \quad &\cr
height10pt && \omit && \omit && \omit && \omit & \cr
&& \quad $Z \to Q+\Qbar+\QQbarsinglet$ \quad &&
\quad $7.7\times10^{-5}$\quad && \quad $3.7 \times 10^{-5}$ \quad &&
\quad $1.6 \times 10^{-5}$ \quad &\cr
height10pt && \omit && \omit && \omit && \omit & \cr
&& \quad $Z \to Q+\Qbar+\QQbaroctet$ \quad &&
\quad $1.1 \times10^{-7}$ \quad && \quad $3.2 \times 10^{-8}$ \quad &&
\quad $5.7 \times 10^{-8}$ \quad &\cr
height10pt && \omit && \omit && \omit && \omit & \cr
&& \quad $Z \to q+\qbar+\QQbaroctet$ \quad &&
\quad $3.3 \times 10^{-4}$ \quad && \quad $9.6 \times 10^{-5}$ \quad &&
\quad $4.1 \times 10^{-5}$ \quad &\cr
height10pt && \omit && \omit && \omit && \omit &\cr
\tablerule}} $$
\centerline{Table I}
\medskip\noindent

$$ \vbox{\offinterlineskip
\def\tablerule{\noalign{\hrule}}
\hrule
\halign {\vrule#& \strut#&
\ \hfil#\hfil & \vrule#&
\ \hfil#\hfil & \vrule#&
\ \hfil#\hfil & \vrule#&
\ \hfil#\hfil & \vrule#&
\ \hfil#\hfil & \vrule#&
\ \hfil#\hfil & \vrule# \cr
\tablerule%
height10pt && \omit && \omit && \omit && \omit && \omit && \omit &\cr
&& Production && QCD && Short Distance && Color && Flavor && Long Distance &\cr
&& Mechanism && Order && Factor && Factor && Factor && Matrix Element &\cr
height10pt && \omit && \omit && \omit && \omit && \omit && \omit &\cr
\tablerule
height10pt && \omit && \omit && \omit && \omit && \omit && \omit &\cr
&& \quad $Z \to g+\QQbaroctet$ \quad &&
\quad $\aS$ \quad && \quad 1 \quad && \quad $\displaystyle{\Nc^2-1 \over 2}$
\quad && \quad 1 \quad && \quad $\octetelement$ \quad &\cr
height10pt && \omit && \omit && \omit && \omit && \omit && \omit &\cr
&& \quad $Z \to Q+\Qbar+\QQbarsinglet$ \quad &&
\quad $\aS^2$ \quad && \quad $\displaystyle{1 \over \d^2}$ \quad &&
\quad $\Bigl( \displaystyle{\Nc^2-1 \over 2\Nc} \Bigr)^2$ \quad &&
\quad 1 \quad && \quad $\singletelement$ \quad &\cr
height10pt && \omit && \omit && \omit && \omit && \omit && \omit &\cr
&& \quad $Z \to Q+\Qbar+\QQbaroctet$ \quad &&
\quad $\aS^2$ \quad && \quad $\displaystyle{1 \over \d^2}$ \quad &&
\quad $\displaystyle{\Nc^2-1 \over 4\Nc^2}$ \quad &&
\quad 1 \quad && \quad $\octetelement$ \quad &\cr
height10pt && \omit && \omit && \omit && \omit && \omit && \omit &\cr
&& \quad $Z \to q+\qbar+\QQbaroctet$ \quad &&
\quad $\aS^2$ \quad && \quad $\displaystyle{(\log \d^2)^2 \over \d^2}$ \quad &&
\quad $\displaystyle{\Nc^2-1 \over 2}$ \quad &&
\quad 5 \quad && \quad $\octetelement$ \quad &\cr
height10pt && \omit && \omit && \omit && \omit && \omit && \omit &\cr
\tablerule}} $$
\centerline{Table II}
\medskip\noindent

\listrefs
\listfigs

\bye